\documentclass[conference]{IEEEtran}
\IEEEoverridecommandlockouts
\usepackage{cite}
\usepackage{amsmath,amssymb,amsfonts}
\usepackage{algorithmic}
\usepackage{graphicx}
\usepackage{textcomp}
\usepackage{xcolor}
\usepackage{cite}
\usepackage{multirow}
\usepackage{booktabs}

\usepackage{url}
\def\BibTeX{{\rm B\kern-.05em{\sc i\kern-.025em b}\kern-.08em
    T\kern-.1667em\lower.7ex\hbox{E}\kern-.125emX}}
\begin{document}

\title{Continual Pre-training for Codec-Based Speech LLMs: Balancing Understanding and Generation
\thanks{The work is done during Internship.}
}

\author{Jiatong Shi$^{1}$ \quad Chunlei Zhang$^{2}$ \quad Jinchuan Tian$^{1}$ \quad Junrui Ni$^{3}$ \quad Hao Zhang$^{2}$ Shinji Watanabe$^{1}$ \quad Dong Yu$^{2}$ \\
$^{1}$ CMU, $^{2}$ Tencent AI Lab, $^{3}$ UIUC}

% \author{Reserved for Double-blind Review}

% \and
% \and
% \and
% \and
% \and

\maketitle

\begin{abstract}
Recent advances in speech language models (LLMs) have extended textual LLMs to the speech domain, but balancing speech understanding and generation remains challenging, especially with codec-based representations. We propose a continual pre-training (CPT) framework that adapts a textual LLM to handle codec-discretized speech, mitigating modality mismatch and preserving linguistic reasoning. Our unified model supports both understanding and generation, achieving strong results across ASR, TTS, S2T-Trans, and S2S-Trans. Notably, we present the first end-to-end, single-pass S2S-Trans system using only neural codec tokens, without intermediate transcriptions, translations, or semantic tokens. CPT proves essential for cross-modal alignment and task generalization, making it a powerful tool for building robust, unified speech LLMs.
\end{abstract}


\section{Introduction}
\label{sec: intro}

Large language models (LLMs) have transformed natural language processing with strong generalization and transfer capabilities across a wide range of tasks~\cite{LLMSurvey, achiam2023gpt, dubey2024llama, yang2024qwen2}. Their autoregressive architecture and unified token interface allow efficient modeling of diverse text-based tasks using a single model. Motivated by this success, recent research has explored extending LLMs to the speech domain, yielding models capable of speech understanding~\cite{kharitonov2022text, gong2023joint, chang2024speechprompt, maiti2024voxtlm} or generation~\cite{wang2023neural, defossez2024moshi, kim2024clamtts, wu2025towards, casanova2025low, guo2024lscodec, guo2025speaking}.

A key obstacle in building balanced speech LLMs is the modality gap between symbolic text and continuous speech. Codec-based speech representations offer high-fidelity generation by preserving low-level acoustic detail~\cite{wang2023neural, yanguniaudio}, yet struggle with semantic tasks like automatic speech recognition~(ASR) due to their lack of abstraction~\cite{shi2024espnetcodeccomprehensivetrainingevaluation, ye2025codec}. Moreover, when such representations are introduced into textual LLMs, the shift in input distribution often causes catastrophic forgetting of language capabilities, undermining the foundation of the model.

To address this, we propose a continual pre-training (CPT) framework that bridges the speech-text modality gap by adapting a pre-trained LLM using codec-discretized speech data. By aligning speech and text in a shared embedding space and training with a unified autoregressive objective, CPT allows the model to learn speech patterns without losing its linguistic competence. We study two CPT configurations: speech-only and joint speech-text, each revealing different tradeoffs in generation and understanding.

We conduct extensive evaluations across four speech tasks: ASR, text-to-speech~(TTS), speech-to-text translation (S2T-Trans), and speech-to-speech translation (S2S-Trans). CPT consistently improves performance in both speech generation and understanding. Speech-only CPT favors generation, while joint CPT preserves textual reasoning. Crucially, we demonstrate for the first time that a single model can perform end-to-end speech-to-\textbf{speech} translation using only codec tokens, without intermediate text or semantic guidance, highlighting the power of CPT for unified speech modeling.

Our key contributions are:
\begin{itemize}
    \item We introduce CPT as an effective strategy to bridge modality mismatch between text and codec-based speech in LLMs, mitigating catastrophic forgetting.
    \item We design and implement a unified architecture for autoregressive speech and text modeling, using shared embeddings and parallel prediction heads.
    \item We demonstrate strong performance across diverse speech tasks, and achieve the first single-pass, intermediate-free S2S-Trans system using only codec tokens.
\end{itemize}

\begin{figure*}[t]
    \centering
    \includegraphics[width=0.85\linewidth]{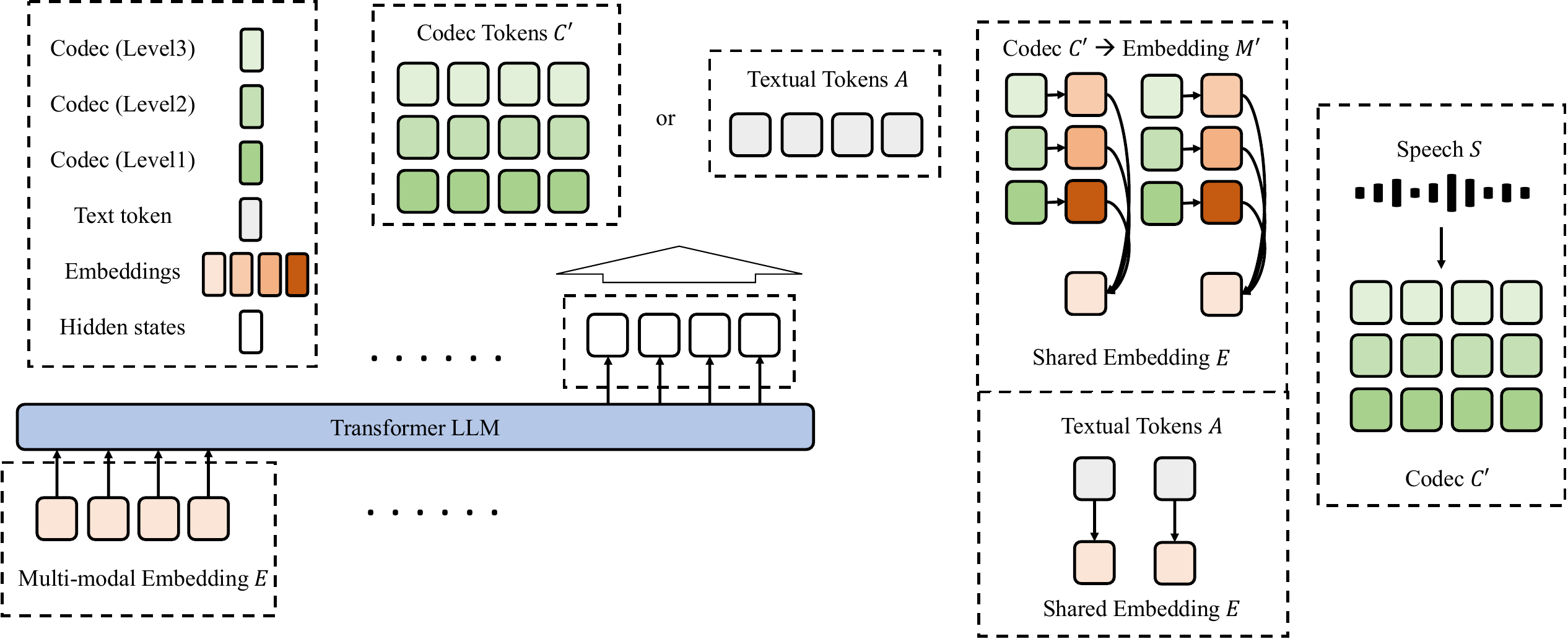}
    \vspace{-9pt}
    \caption{The architecture of the codec-based speech LLM. Codec tokens and textual tokens are converted into multi-modal shared embeddings, as shown on the right side of the figure. These shared embeddings are then fed into a Transformer-based LLM, which features parallel prediction heads designed for predicting either codec tokens or textual tokens. Details are discussed in Sec.~\ref{sec: method}.}
    \label{fig:codec speech llm}
    \vspace{-10pt}
\end{figure*}

\section{Related Works}
\label{sec: related works}

\noindent \textbf{Speech Language Modeling}.  
Speech LLMs commonly extend textual LLMs by introducing discrete tokenizations of speech to unify modalities. SSL-based tokens, used in models like SpeechGPT~\cite{zhang-etal-2023-speechgpt, rubenstein2023audiopalm, maiti2024voxtlm, wu2024speechcomposerunifyingmultiplespeech}, support semantic-level understanding and align well with text token spaces. However, due to their lossy nature and reliance on vocoders, they often underperform in generation, particularly in multi-speaker synthesis~\cite{lee-etal-2022-direct, maiti2024voxtlm}. To improve speech generation, codec-based tokens have been adopted in models like Moshi~\cite{defossez2024moshi}, Neural Codec LM~\cite{wang2023neural}, ClamTTS~\cite{kim2024clamtts}, and UniAudio~\cite{yanguniaudio}, offering high-fidelity reconstruction by preserving acoustic detail. However, codec tokens present challenges in understanding tasks such as ASR due to a modality gap~\cite{shi2024espnetcodeccomprehensivetrainingevaluation, chang2024exploring, Dhawan_2024}.

Our work departs from models like SpeechGPT~\cite{zhang2023speechgpt}, which use SSL tokens better suited for comprehension, and from \cite{zengscaling}, which employs Whisper-derived tokens that are semantically rich but limited in acoustic representation. In contrast, our approach leverages codec tokens that retain low-level speech information, enabling both high-quality generation and generalization to understanding tasks. We also differ from Moshi~\cite{defossez2024moshi}, which uses a two-stage decoding process, generating text tokens first, then predicting codec tokens conditioned on them. This requires intermediate alignment (i.e., force-alignment). Our model avoids such dependency by using interleaving speech and text tokens, simplifying training and supporting direct speech-to-speech modeling. Mini-Omni~\cite{xie2024mini} uses a pre-trained speech encoder for input and a multi-stream codec for output, projecting speech into the text domain via additional adapters. Despite being end-to-end trainable, the system fundamentally relies on a speech-to-text-to-speech pipeline. In contrast, our approach maintains speech and text within a unified token space, avoiding explicit cross-modal conversions and offering a more integrated framework.

\noindent \textbf{Continual Pre-training (CPT)}.  
CPT has shown promise for adapting LLMs to new modalities or domains~\cite{wu2024continuallearninglargelanguage}. In speech LLMs, modality mismatch between text and speech (e.g., sequence length, acoustic information) often causes catastrophic forgetting~\cite{zhan2024anygpt, huang2024dynamic}. Prior CPT efforts using SSL tokens~\cite{zhang2023speechgpt, rubenstein2023audiopalm} effectively extended textual LLMs but suffered from poor synthesis quality. In contrast, neural codecs offer better reconstruction fidelity~\cite{wu2024codecsuperbindepthanalysissound, kim2024neuralspeechaudiocoding}, though the multi-stream nature complicates integration with text-based LLMs~\cite{chen2024vall, NEURIPS2023_94b472a1}.

While CPT has been increasingly adopted in multimodal LLM adaptation, our work is one of the first to apply it for codec-based speech LLMs, showing its practical benefits in balancing generation and understanding.

\noindent \textbf{Remarks on Concurrent Works with CPT}. Several concurrent efforts have recently proposed end-to-end speech language models that directly operate on speech tokens and adopt CPT or related adaptation strategies. Notable examples include ESPnet-SpeechLM~\cite{tian2025espnet}, Qwen2.5-Omni~\cite{xu2025qwen2}, and Kimi-Audio~\cite{ding2025kimi}, which integrate codec-based representations into LLM backbones and apply large-scale speech pre-training for downstream speech tasks. While these models leverage CPT-related mechanisms to incorporate speech modalities, they do not explicitly study the role or effect of CPT in isolation. In contrast, our work provides a detailed and controlled examination of CPT, including speech-only versus joint speech-text configurations, and investigates its impact on both understanding and generation tasks. We show that CPT not only improves performance but is essential for enabling single-pass S2S-Trans in a unified framework. This analysis contributes new insights into modality alignment and catastrophic forgetting, which remain under-explored in these concurrent works.

\noindent \textbf{S2S-Trans}. Most prior S2S-Trans systems follow a multi-stage pipeline, decomposing the task into ASR, machine translation (MT), and TTS modules~\cite{ahmad2024findings, agarwal2023findings}. These cascaded systems rely heavily on intermediate transcriptions, translations, or semantic representations, and typically require modality alignment and modular training. Even recent end-to-end efforts like AudioPaLM~\cite{rubenstein2023audiopalm} and Seamless~\cite{barrault2023seamless} use intermediate supervision or decoding stages tied to text.

In contrast, our model performs S2S-Trans without using intermediate transcriptions, translations, or semantic tokens at any stage, relying solely on codec tokens throughout. To our knowledge, this is the first demonstration of a single-pass, decoder-only speech LLM that enables S2S-Trans using only codec tokens and no intermediate supervision

\section{Methodology}
\label{sec: method}

This section outlines the key components of the proposed method, including the speech tokenizer, model architecture, and the training strategy employed for CPT.

\noindent \textbf{Speech Tokenizer}. Following Sec.~\ref{sec: related works}, we use neural codecs to discretize speech into codec tokens. The tokenizer consists of an encoder, a quantizer with $L$ codebooks $\{\mathcal{B}^i\}_{i=1}^L$, and a decoder. Given a speech signal $S \in \mathbb{R}^{1 \times T_S}$, the encoder produces hidden states $Q \in \mathbb{R}^{D^\text{embed} \times T_Q}$. The quantizer maps $Q$ into discrete codes $C \in (\mathcal{B}^1 \times \dots \times \mathcal{B}^L)^{T_C}$ and reconstructs embeddings $E \in \mathbb{R}^{D^\text{embed} \times T_C}$ via learned codebook embeddings $M$. Here $T_S$, $T_Q$, and $T_C$ are lengths of speech $S$, hidden states $Q$, and codecs $C$, respectively.

These embeddings are then passed to the decoder to reconstruct the original signal, $\hat{S} = \mathrm{Decoder}(E)$. The discrete codes $C$ serve as the I/O representation for our codec-based speech LLM.

\noindent \textbf{Model Architecture}. The overall architecture is shown in Figure~\ref{fig:codec speech llm}. Both speech codec tokens and text tokens are projected into a shared multimodal embedding space. For speech, we select $L' (\leq L)$ codec streams from $C$ to form $C' \in (\mathcal{B}^1 \times \dots \times \mathcal{B}^{L'})^{T_C}$, which are embedded and summed into $E^\text{speech} \in \mathbb{R}^{D^\text{embed} \times T_C}$. Text tokens $A \in \mathcal{V}^{T_A}$ are directly embedded as $E^\text{text} \in \mathbb{R}^{D^\text{embed} \times T_A}$, where $T_A$ is the length of text. These interleaved embeddings $E$ are processed by a decoder-only Transformer. For simplicity, we denote the combined token sequence as $R \in (\mathcal{B}^{1} \times \dots \times \mathcal{B}^{L'} \times \mathcal{V})^{T_R}$, where $T_R$ is the combined sequence length.

The Transformer backbone is adapted from Qwen1.5~\cite{yang2024qwen2}. We retain the architecture, tokenizer, and text embeddings of the original model to preserve language capabilities. The output head for text is partially initialized from Qwen1.5, while the codec-related vocabulary is randomly initialized. Additional heads are added to predict each codec level \textit{in parallel}, avoiding hierarchical dependency modeling. Compared to multi-stage or multi-scale alternatives~\cite{chen2024vall, NEURIPS2023_94b472a1, yanguniaudio}, our design remains simple and efficient.

\noindent \textbf{Model Training}. The formulation discussed in the previous two sections provides a unified interface for speech and text modalities within a single end-to-end model, where speech is represented in multiple streams, text in a single stream, with interleaving embedding frames.

However, this formulation has limitations when incorporating a pre-trained textual LLM due to significant modality mismatches, particularly with speech codecs. To address this, we propose using CPT to align modalities, enabling a unified model for both understanding and generation. This approach leverages the strong foundation of textual LLMs while maintaining high-fidelity speech acoustics through advanced codecs.

In the original textual LLM, the pre-training objective maximizes the likelihood of next-token prediction over text tokens:
\begin{equation}
\label{eq: text-pt}
    \theta_{\text{PT-}A} = \mathrm{argmax}_\theta \prod_t^{T_A} P_\theta(a_t | a_{<t}),
\end{equation}
where $a_{<t}$ denotes the preceding text tokens. Under CPT with speech-only data, the objective adopts the same form but over speech codec tokens:
\begin{equation}
\label{eq: audio-pt}
    \theta_{\text{CPT-}C} = \mathrm{argmax}_\theta \prod_t^{T_C} P_\theta(c_t | c_{<t}).
\end{equation}

The purpose of CPT is to inject knowledge about speech modality into the LLM transformer base. Therefore, we conduct two pre-training configurations: one with speech data only following Eq.~\eqref{eq: audio-pt}, and the other with bi-modal (i.e., speech and text) datasets. Details of data preparation are discussed in Sec.~\ref{ssec: dataset}.  We denote a sequence of speech and text tokens $R = (r_1, r_2, ..., r_{T_R})$ with a length of $T_R$. $r_t \in (\mathcal{B}^1 \times \mathcal{B}^2 \times ... \times \mathcal{B}^{L'} \times \mathcal{V})$ can represent either a speech codec token $(c^1_t, c^2_t, ..., c^{L'}_t, \text{null})$ or text token $(\text{null}, ..., \text{null}, a_t)$. The pre-training objective follows the next-token prediction task:
\begin{equation}
\label{eq: pt}
    \theta_\text{CPT} = \mathrm{argmax}_\theta \prod_t^{T_R} P_\theta(r_t | r_{<t}),
\end{equation}
where $\theta$ is the model illustrated in Fig.~\ref{fig:codec speech llm} and $r_0$ is the start of sentence. 

Following the CPT stage, we conduct SFT where a prompt token sequence $R^{\text{inp}}$ (a composed text/speech token sequences) is first processed before the target sequence $R^{tgt}$. The objective for the SFT is:
\begin{equation}
\label{eq: ft}
    \theta_\text{SFT} = \mathrm{argmax}_\theta P_\theta(R^{\text{tgt}} | R^{\text{inp}}).
\end{equation}

\section{Experimental Setup}

In this section, we detail the experimental settings for CPT and SFT to evaluate the effectiveness of the proposed method across various downstream tasks. Specifically, we compare CPT-ed models with those that are either randomly initialized or initialized from a textual LLM.

The basic speech tokenizer follows the SoundStream architecture~\cite{Zeghidour2021SoundStreamAE}. Instead of using a complex short-term Fourier transform (STFT) discriminator, we adopt the discriminators from \cite{Kumar2023HighFidelityAC}, which include a multi-frequency STFT discriminator, a multi-scale discriminator, and a multi-period discriminator.

\subsection{Pre-training Data Preparation}
\label{ssec: dataset}
The pre-training data consists of around 140k hours of English and Mandarin speech, along with corresponding transcriptions and translations.\footnote{CPT uses over 140k hours of English and Chinese speech from public corpora such as Aishell~\cite{bu2017aishell, du2018aishell, shi21c_interspeech}, Wenetspeech~\cite{zhang2022wenetspeech}, Gigaspeech~\cite{chen21o_interspeech}, Librispeech~\cite{panayotov2015librispeech}, MLS~\cite{pratap20_interspeech}, TEDLIUM3~\cite{hernandez2018ted}, WSJ~\cite{rottland1997wsj}, and CommonVoice~\cite{ardila-etal-2020-common}, supplemented by in-house Mandarin data and text from Wikipedia, The Pile~\cite{gao2020pile}, and subtitle corpora.} For training consistency on punctuation, we use a BERT-based punctuation restoration model\footnote{\url{https://huggingface.co/felflare/bert-restore-punctuation}} to recover the punctuation for textual data without punctuation (e.g., Librispeech~\cite{panayotov2015librispeech}).

Most of the corpora included were originally designed for ASR or TTS purposes and did not include translations in their official release, which may prevent full alignment of the semantic spaces between the two languages. To semantically align English and Mandarin in the speech LLM, we supplement the data using an internal machine translation model to generate translations between English and Mandarin (EN-to-ZH and ZH-to-EN) based on the original speech transcriptions. For each utterance in the pre-training dataset, we use the paired tuple (i.e., speech, transcription, and translation) to create six tasks: (1)~speech continuation, which predicts future speech tokens based on previous segments; (2)~language modeling, which generates the following textual tokens from a given context; (3)~ASR, which transcribes speech into text; (4)~TTS, which generates speech from text using a target speaker's speech segment as a prompt; (5)~speech-to-text translation~(S2T-Trans), which translates speech in one language to text in another; and (6)~text-to-speech translation~(T2ST), which generates translated speech from a source text using a target speaker's prompt. 

Importantly, \textit{we did not include S2S-Trans during continual pre-training}. This allows us to treat S2S-Trans as a true unseen downstream task and evaluate whether the CPT framework successfully bridges source and target languages in the speech modality, without relying on explicit intermediate information such as transcriptions, translations, or semantic tokens. 

Based on the six tasks mentioned, the final sequence is formulated using the pattern ``\texttt{(Condition)(Prompt)(Target)}''. Additionally, a boundary token is inserted for each speech or text segment to mark the start or end of the segment.

\subsection{Continual Pre-training}

We adopt \texttt{Megatron-LM}\footnote{\url{https://github.com/NVIDIA/Megatron-LM}} with tensor parallelism as our training framework \cite{shoeybi2019megatron}, using Qwen1.5-0.5B as the base pre-trained textual LLM \cite{yang2024qwen2}.

We carry out two CPT experiments: one using only speech data, and the other incorporating both speech and text modalities. For the mixed-modality training, we aim to balance different domains and tasks by constructing a data loader that samples data from various tasks. Specifically, the loader randomly selects data for each of the six speech-related tasks with a 15\% probability. To maintain the reasoning capacity of the original textual LLM and avoid catastrophic forgetting, we include text-only data for the remaining 10\%. Of this textual data, 5\% comes from general domains such as books, YouTube titles, and Wikipedia, while the other 5\% is sourced from an in-house machine translation~(MT) corpus to enhance the model's translation capabilities. This design aligns with previous CPT approaches in both textual and multimodal LLMs, where even limited text-based pre-training acts as a stabilizing regularizer, anchoring the model’s internal representations to those learned during pre-training~\cite{zhai2023investigating, sun2020ernie}.

Pre-training is performed with tensor parallelism of 8, enabling large-batch training~\cite{korthikanti2023reducing}. We use a global batch size of 640, a sequence length of 4,096, and train for 40k steps with gradient clipping set to 1.0 and normalization $\epsilon=10^{-5}$. The model, with 943.5M parameters, is trained using \texttt{BFloat16} and the distributed AdamW optimizer (peak LR: $1\text{e}{-5}$; min LR: $1\text{e}{-6}$). The vocabulary is expanded to 155,012 (including padding and 293 shape-adjustment tokens) to support parallelism.

Additionally, to identify the effect of CPT, we conduct experiments with a random-initialized 0.5B model that has the same architecture as Qwen1.5-0.5B and the pre-trained Qwen1.5-0.5B for comparison.\footnote{They are denoted as ``No Initialization" and ``Text LLM Initialization" in the following discussion.}

\subsection{Downstream Tasks}
\label{ssec: downstream}

As formulated in Eq.~\eqref{eq: ft}, the downstream evaluation is performed by fine-tuning the pre-trained model across four tasks: ASR, TTS, S2T-Trans, and S2S-Trans. ASR, TTS, and S2T-Trans are used to assess the model's basic speech understanding and generation abilities. S2S-Trans can be broken down into ASR, MT, and TTS, making it a more comprehensive task that evaluates both understanding and generation capabilities.

All four fine-tuning tasks follow the sequence formulation discussed in Sec.~\ref{ssec: dataset}, with different \texttt{(Condition)}, \texttt{(Prompt)}, and \texttt{(Target)}, related to the tasks. Notably, the task prompts are also generated following the pipeline in Sec.~\ref{ssec: dataset} but they do not overlap with the pre-training prompts.

\noindent \textbf{ASR}: For the ASR task, we evaluate the model using the Librispeech dataset for English and the Aishell2 dataset for Mandarin \cite{panayotov2015librispeech, du2018aishell}. The \texttt{(Condition)} consists of a sequence of speech codec tokens, followed by a natural language prompt. As described in Sec.~\ref{ssec: dataset}, the \texttt{(Target)} sequence is the transcription with restored punctuation. To further improve model performance, we apply random time-domain masking, similar to the approach in \cite{chang23b_interspeech, chang2024exploring}.

During inference, the model autoregressively predicts the transcription by feeding the \texttt{(Condition)(Prompt)} sequence into the system. To enhance decoding performance, we employ beam search, using a beam size of 8. We measure word error rate (WER) for English ASR and character error rate (CER) for Mandarin ASR.

\noindent \textbf{TTS}: We focus on the multi-speaker TTS task with LibriTTS~\cite{zen19_interspeech}. For the task formulation, we follow VaLL-E style input where the condition includes a three-second speaker prompt and the text~\cite{wang2023neural}. For the target text, we restore the punctuation similar to the ASR task.

Since greedy search tends to produce trivial outputs, we adopt a sampling-based inference with a top-$k$ strategy, setting $k=30$, as used in prior works \cite{wang2023neural, yanguniaudio, tian2024preference}. To further increase the diversity of generated speech, we re-scale the predicted logits using a temperature of 1.5.

For evaluation, we use: WER from a pre-trained ASR model, speaker similarity~(SPK-SIM) from a pre-trained speaker embedding model, and an automatic speech quality predictor based on a pre-trained mean opinion score (MOS) predictor. Specifically, we use Whisper-large-V3 \cite{radford2023robust} for WER evaluation, a pre-trained Rawnet3 model \cite{jung24c_interspeech} trained on Voxceleb for speaker embedding extraction, and UTMOS \cite{saeki2022utmos} as the MOS predictor. Following common practice in previous works \cite{wang2023neural, yanguniaudio, he2024emilia, tian2024preference}, we generate five samples per test instance using the sampling strategy and report the average score across each metric.

\noindent \textbf{S2T-Trans}: The S2T-Trans adopts a task formulation as the ASR task by simply replacing the input speech with source language speech and target transcription with target language translation. The prompts are changed to task-related prompts accordingly. We test two corpora: CoVOST2 and GigaST~\cite{wang21s_interspeech, ye23b_interspeech}. For CoVOST2, we focus on two translation directions, including English-to-Mandarin~(EN-to-ZH) and Mandarin-to-English~(ZH-to-EN). For GigaST, we only focus on EN-to-ZH. We use SacreBLEU to evaluate with BLEU score~\cite{post-2018-call}.

\noindent \textbf{S2S-Trans}: We conduct S2S-Trans using the GigaS2S corpus, which supplements the GigaST corpus with a single-speaker TTS model~\cite{ye23b_interspeech}.\footnote{\url{https://github.com/SpeechTranslation/GigaS2S}} Due to data constraints, we focus only on the English-to-Mandarin~(EN-to-ZH) translation direction. Since the target speech is single-speaker, the \texttt{(Condition)} consists solely of source language speech. Crucially, \textit{this task is not included in the continual pre-training stage}, no paired speech-to-speech training data or pseudo-targets were seen during CPT. This design ensures that S2S-Trans serves as a held-out, unseen task to evaluate the model’s ability to generalize across modalities. Furthermore, no intermediate transcriptions, translations, or semantic representations are used during either CPT or SFT. The model must learn to directly map input speech in one language to output speech in another, purely through codec token modeling. Prompt generation follows the same pattern as other tasks, but S2S-Trans remains the task evaluated entirely without prior exposure.

For inference, we use the same top-$k$ strategy employed in the TTS task. Evaluation metrics include ASR-BLEU for translation accuracy and UTMOS for speech quality.\footnote{Although UTMOS was trained on English speech, which might introduce some language mismatch in the scoring, prior work \cite{huang22f_interspeech} has shown that UTMOS still achieved reasonable correlation scores when evaluating out-of-domain Chinese speech. Thus, we continue to use it for our speech quality evaluation.}

\section{Results and Discussion}
\label{ssec: result}

\subsection{S2S-Trans Experiments}

% S2S-Trans represents the most challenging and novel task in our evaluation. The goal is to generate speech in a target language directly from input speech in a source language, without relying on intermediate transcriptions, translations, or semantic tokens. As shown in Table~\ref{table:S2S-Trans}, only the CPT models successfully complete this task. The model trained with joint speech-text CPT achieves the best performance, with strong results in both ASR-BLEU and UTMOS. The speech-only CPT model also performs well, indicating that training on raw speech tokens helps the model develop generative capability. However, the joint CPT model is more effective at learning the cross-lingual mapping required for this task. In contrast, models trained without CPT fail to converge, even after careful hyperparameter tuning. This confirms that CPT plays an essential role in aligning modalities and enabling generalization to new speech tasks.

The speech-to-speech translation (S2S-Trans) task serves as the primary evaluation benchmark in this paper, designed to rigorously test the generalization ability of our continual pre-training (CPT) framework. Unlike other downstream tasks that mirror the tasks used during CPT (e.g., ASR, TTS), S2S-Trans was deliberately excluded from the pre-training stage (see Sec.~\ref{ssec: dataset}). This means the model has never seen a speech-to-speech objective or paired source-target speech inputs during CPT, nor has it been exposed to intermediate textual supervision like transcriptions or translations.

The motivation behind this task is to determine whether CPT can improve a codec-based speech LLM with emergent cross-lingual speech translation ability, purely through exposure to related but decoupled tasks, such as ASR, TTS, and S2T-Trans, without explicitly training on S2S mappings. A successful result would demonstrate that CPT is not merely beneficial for fine-tuning stability, but is in fact crucial for enabling complex, unseen modality transfers in a zero-shot-like setting.

As shown in Table~\ref{table:S2S-Trans}, only models trained with CPT are able to perform the task, and the joint speech-text CPT variant achieves the best performance in both ASR-BLEU and UTMOS. The speech-only CPT variant also succeeds, validating that exposure to raw speech token prediction improves generative capacity. In stark contrast, models without CPT fail to converge on this task, confirming that CPT is essential for aligning modalities and enabling end-to-end speech-to-speech translation in a unified codec-token space.

We also compare to prior systems, including HW-TSC, which uses a cascaded ASR-MT-TTS pipeline, and Vec-Tok, which relies on additional supervision
~\cite{zhu2023vec, wu2024improving}. While these baselines use intermediate representations or modular components, our model performs comparably using a single unified decoder-only LLM trained without such supervision. This result provides the first demonstration of end-to-end, single-pass S2S-Trans using only codec tokens in a language model framework.

\begin{table}[t]
\centering
\caption{S2ST performance on GigaST. \textsuperscript{$\dagger$} indicates that the ASR-BLEU scores were calculated using different ASR systems, as described in Sec.~\ref{ssec: result}.}
\vspace{-5pt}
\resizebox{0.7\linewidth}{!}{
\begin{tabular}{l|c|c}
\toprule
\multirow{2}{*}{Pre-training} & \multicolumn{2}{|c}{GigaST}  \\
\cmidrule{2-3} 
 & ASR-BLEU & UTMOS  \\
 \midrule
 Vec-Tok\textsuperscript{$\dagger$}~\cite{zhu2023vec} & 21.6 & -   \\
 HW-TSC\textsuperscript{$\dagger$}~\cite{wu2024improving} & \textbf{33.6} & - \\
 \midrule
 Speech CPT & 28.0 & 3.41 \\
 Speech \& Text CPT & 33.4 & \textbf{3.66} \\

\bottomrule
\end{tabular}
}
% \vspace{-15pt}
\label{table:S2S-Trans}
\end{table}

\subsection{ASR, TTS, and S2T-Trans Experiments}

To complement the S2S-Trans analysis, we evaluate the model on three tasks that were explicitly included during CPT: ASR, TTS, and S2T-Trans. These tasks serve two purposes: (1) verifying that the model maintains strong performance on tasks seen during CPT, and (2) revealing how speech-only vs. joint CPT influences task-specific capabilities.

\begin{table}[t]
\centering
\caption{ASR performance on LibriSpeech and Aishell2. Models marked with \textsuperscript{*} indicate pre-trained models did not undergo continual pre-training. \textsuperscript{+} stands models that do not use neural codecs as their speech representation. We report WER for Librispeech and CER for Aishell2.} \vspace{-5pt}
\resizebox{1.0\linewidth}{!}{
\begin{tabular}{l|c|c|c|c}
\toprule
\multirow{2}{*}{Models} & \multirow{2}{*}{Param.} &  \multicolumn{2}{|c|}{LibriSpeech} & Aishell2 \\
\cmidrule{3-5}
 & & Test-clean & Test-other & Test-overall \\
 \midrule 
 VoxtLM\textsuperscript{+}~\cite{maiti2024voxtlm} & 1B & \textbf{2.7} & 6.5 & - \\
 SALMONN\textsuperscript{+}~\cite{tang2024salmonn} & 7B & 2.1 & 4.9 & - \\
 % \midrule
 AnyGPT~\cite{zhan2024anygpt} & 7B & 8.5 & - & - \\
 Qwen-Audio2~\cite{chu2024qwen2audiotechnicalreport} & 7B & 1.3 & 3.6 & 3.0 \\
 ESPnet-SpeechLM~\cite{tian2025espnet} & 2B & 2.8 & 5.9 & - \\
 \midrule
 No Initialization\textsuperscript{*} &\multirow{4}{*}{1B} & 5.5 & 9.5 & 15.5 \\
 Text LLM Initialization\textsuperscript{*} &  & 4.8 & 8.5 & 13.1 \\
 Speech CPT &  & 5.5 & 8.9 & 13.0 \\
 Speech \& Text CPT &  & 3.7 & \textbf{6.3} & \textbf{7.2} \\

\bottomrule
\end{tabular}
}
\vspace{-15pt}
\label{table:asr}
\end{table}
\noindent \textbf{ASR}. Table~\ref{table:asr} shows that the model with joint speech-text CPT achieves the best performance among our variants, confirming the importance of mixed-modality CPT in preserving linguistic competence. In contrast, the speech-only CPT and non-CPT variants yield higher WERs, especially on Librispeech, suggesting that speech continuation alone is insufficient for strong transcription performance.

We also compare against a range of existing speech LLMs. Models such as VoxtLM~\cite{maiti2024voxtlm}, SALMONN~\cite{tang2024salmonn}, and Qwen-Audio2~\cite{chu2024qwen2audiotechnicalreport} achieve lower WERs, particularly on test-clean. However, these models differ substantially in design and training objectives. Specifically, SALMONN and Qwen-Audio2 leverage dedicated audio encoders, which are optimized for speech understanding but not suitable for generation. Similarly, VoxtLM and ESPnet-SpeechLM~\cite{tian2025espnet} incorporate SSL-derived semantic tokens, which provide high-level linguistic abstraction but are known to degrade speech synthesis quality~\cite{yanguniaudio}.

In contrast, our model operates purely on codec tokens and supports both understanding and high-fidelity generation in a single decoder-only architecture. While the joint CPT model underperforms SSL-based systems on clean speech, it remains competitive on the more challenging test-other set and on Mandarin (Aishell2), demonstrating robust generalization in realistic conditions.

\noindent \textbf{TTS}. As shown in Table~\ref{table:tts}, the speech-only CPT model performs best, consistent with its exposure to speech continuation tasks. Joint CPT achieves comparable intelligibility (WER) and speaker similarity, though slightly lower MOS. Both CPT variants clearly outperform non-CPT baselines, confirming that CPT improves speech generation capabilities.

\begin{table}[t]
\centering
\caption{TTS performance on LibriTTS. Models marked with \textsuperscript{*} indicate pre-trained models that did not undergo continual pre-training. \textsuperscript{$\circ$} corresponds to a version trained on LibriTTS.} \vspace{-5pt}
\resizebox{0.8\linewidth}{!}{
\begin{tabular}{l|c|c|c}
\toprule
Models & UTMOS & WER & SPK-SIM \\
\midrule
 UniAudio\textsuperscript{$\circ$}~\cite{yanguniaudio} & 3.64 & 13.1 & 0.43 \\
 Vall-E\textsuperscript{$\circ$}~\cite{chen2024vall} & 3.29 & 5.4 & 0.63 \\
 Delay-TTS\textsuperscript{$\circ$}~\cite{copet2023simple} & 3.44 & 6.8 & 0.54 \\
 \midrule
 No Initialization\textsuperscript{*} &  3.01 & 17.5 & 0.55 \\
Text LLM Initialization\textsuperscript{*} &  2.78 & 18.8 & 0.51 \\
 Speech CPT &  \textbf{3.65} & \textbf{3.7} & \textbf{0.66} \\
 Speech \& Text CPT &  3.59 & \textbf{3.7} & 0.65 \\

\bottomrule
\end{tabular}
}
\vspace{-10pt}
\label{table:tts}
\end{table}
\begin{table}[t]
\centering
\caption{S2TT performance on CoVOST2 and GigaST. The performance is reported in BLEU. Models marked with \textsuperscript{*} indicate pre-trained models that did not undergo continual pre-training. \textsuperscript{$\bullet$} stands that external machine translation data is used.} \vspace{-5pt}
\resizebox{\linewidth}{!}{
\begin{tabular}{l|c|c|c}
\toprule
\multirow{2}{*}{Pre-training} & \multicolumn{2}{|c|}{CoVoST2} & GigaST \\
\cmidrule{2-4} 
 & EN-to-ZH & ZH-to-EN & EN-to-ZH  \\
 \midrule
 Fairseq ST~\cite{wang21s_interspeech} & 25.4 & 5.8 & -  \\
 OWSM-v3~\cite{peng2023reproducing} & \textbf{33.4} & 13.6 & -  \\
 GigaST\textsuperscript{$\bullet$}~\cite{ye23b_interspeech} & - & - & 38.0 \\
 LLM-ST\textsuperscript{$\bullet$}~\cite{huang2023speech} & - & - & \textbf{39.6} \\
 \midrule
 No Initialization\textsuperscript{*} & 25.5 & 5.8 & 30.4 \\
  Text LLM Initialization\textsuperscript{*} & 28.9 & 9.9 & 33.2 \\
 Speech CPT & 24.8 & 5.4 & 33.1 \\
 Speech \& Text CPT & 33.1 & \textbf{16.1} & 37.5 \\

\bottomrule
\end{tabular}
}
\vspace{-15pt}
\label{table:S2T-Trans}
\end{table}

We also benchmark against a TTS-specialized UniAudio model~\cite{yanguniaudio}, using the same tokenizer. Our CPT-ed models match or exceed its MOS and greatly outperform it in WER and SPK-SIM. These results demonstrate that CPT not only brings codec-based speech LLMs closer to specialized models but also enables  consistent multi-task performance.

\noindent \textbf{S2T-Trans}. Table~\ref{table:S2T-Trans} presents results on CoVOST2 and GigaST. The joint CPT model achieves the highest BLEU scores, confirming strong cross-lingual understanding. Interestingly, the speech-only CPT model performs worse than the no-init baseline, further echoing the ASR findings: speech-only CPT benefits generation but not comprehension.

We also compare with Fairseq-ST and OWSM-v3~\cite{wang21s_interspeech, ye23b_interspeech}. Our joint CPT model matches or outperforms these systems, despite using codec tokens rather than semantic or SSL-based representations. This confirms that CPT enables the proposed model to remain competitive even on tasks designed for traditional S2T models.

\noindent \textbf{Empirical Comparison with Mini-Omni and Moshi}: Table~\ref{tab:miniomni-moshi-comparison} summarizes the reported ASR and TTS results. While direct comparison is affected by differences in training data, tokenization strategies, and evaluation pipelines, our approach demonstrates stronger performance across both ASR and TTS tasks. This supports the efficacy of our joint speech-text CPT strategy.

\begin{table}[t]
    \centering
    \caption{Performance comparison with Mini-Omni and Moshi. ASR WER on LibriSpeech test-clean; TTS WER is computed from a pre-trained ASR model on LibriTTS test-clean.}
    \vspace{-5pt}
    \label{tab:miniomni-moshi-comparison}
    \begin{tabular}{l|c|c}
        \toprule
        \textbf{Model} & \textbf{ASR WER (↓)} & \textbf{TTS WER (↓)} \\
        \midrule
        Mini-Omni~\cite{xie2024mini} & 4.5 & -- \\
        Moshi~\cite{defossez2024moshi} & 5.7 & 4.7 \\
        Ours (Joint Speech/Text CPT) & \textbf{3.7} & \textbf{3.7} \\
        \bottomrule
    \end{tabular}
    \vspace{-10pt}
\end{table}

\begin{table}[t]
    \centering
    \caption{Perplexity on the LAMBADA benchmark~\cite{paperno2016lambada} to assess retention of textual reasoning after CPT. Lower is better.}
    \vspace{-5pt}
    \begin{tabular}{l|c}
    \toprule
    \textbf{Model} & \textbf{Perplexity} ($\downarrow$) \\
    \midrule
    Qwen1.5-0.5B (Text LLM Initialization) & 36.51 \\
    Speech-only CPT & 47.88 \\
    Speech \& Text CPT & 38.59 \\
    \bottomrule
    \end{tabular}
    \label{tab:lambada}
    \vspace{-15pt}
\end{table}

\subsection{Perplexity Analysis on Catastrophic Forgetting}
To further assess catastrophic forgetting, we evaluate the language modeling capability of our CPT-ed models on LAMBADA~\cite{paperno2016lambada}, a benchmark designed to test long-range textual reasoning. As shown in Table~\ref{tab:lambada}, the perplexity of the speech-only CPT model degrades substantially, suggesting a significant loss of the original textual knowledge. In contrast, the joint speech-text CPT model retains most of the linguistic capacity of the base Qwen1.5-0.5B, with only a marginal increase in perplexity. This highlights that a small proportion of text data during CPT plays a crucial role in stabilizing the model and mitigating forgetting, supporting our design choice for mixed-modality CPT.

\section{Conclusion}
We explore CPT as an effective strategy to extend codec-based speech LLMs for speech translation-related tasks. By carefully formulating our pre-training data, we adapt a pre-trained textual LLM in two configurations, one with speech-only data and another with a joint speech-text approach. Our extensive experiments on ASR, TTS, S2T-Trans, and S2S-Trans tasks show that CPT can significantly enhance performance. In particular, speech-only CPT yields notable improvements for TTS, while joint speech-text CPT strikes a balance between understanding and generation, ultimately delivering high-quality end-to-end S2S-Trans. These findings underscore the potential of CPT in addressing issues such as catastrophic forgetting and modality mismatch, thereby advancing the development of robust multimodal language models.\footnote{Some generated audio samples are available at \url{https://hiddenmeprivate.github.io/}}

\bibliographystyle{IEEEbib}
\bibliography{ref}

@article{dubey2024llama,
  title={The {Llama} 3 herd of models},
  author={Dubey, Abhimanyu and Jauhri, Abhinav and Pandey, Abhinav and Kadian, Abhishek and Al-Dahle, Ahmad and Letman, Aiesha and Mathur, Akhil and Schelten, Alan and Yang, Amy and Fan, Angela and others},
  journal={arXiv preprint arXiv:2407.21783},
  year={2024}
}

@article{achiam2023gpt,
  title={{GPT}-4 technical report},
  author={Achiam, Josh and Adler, Steven and Agarwal, Sandhini and Ahmad, Lama and Akkaya, Ilge and Aleman, Florencia Leoni and Almeida, Diogo and Altenschmidt, Janko and Altman, Sam and Anadkat, Shyamal and others},
  journal={arXiv preprint arXiv:2303.08774},
  year={2023}
}

@article{LLMSurvey,
    title={A Survey of Large Language Models},
    author={Zhao, Wayne Xin and Zhou, Kun and Li, Junyi and Tang, Tianyi and Wang, Xiaolei and Hou, Yupeng and Min, Yingqian and Zhang, Beichen and Zhang, Junjie and Dong, Zican and others},
    year={2023},
    journal={arXiv preprint arXiv:2303.18223},
    url={http://arxiv.org/abs/2303.18223}
}

@article{yang2024qwen2,
  title={Qwen2 technical report},
  author={Yang, An and Yang, Baosong and Hui, Binyuan and Zheng, Bo and Yu, Bowen and Zhou, Chang and Li, Chengpeng and Li, Chengyuan and Liu, Dayiheng and Huang, Fei and others},
  journal={arXiv preprint arXiv:2407.10671},
  year={2024}
}

@inproceedings{huang2024dynamic,
  title={Dynamic-{SUPERB}: Towards a dynamic, collaborative, and comprehensive instruction-tuning benchmark for speech},
  author={Huang, Chien-yu and Lu, Ke-Han and Wang, Shih-Heng and Hsiao, Chi-Yuan and Kuan, Chun-Yi and Wu, Haibin and Arora, Siddhant and Chang, Kai-Wei and Shi, Jiatong and Peng, Yifan and others},
  booktitle={Proc. ICASSP},
  pages={12136--12140},
  year={2024},
  organization={IEEE}
}

@inproceedings{kharitonov2022text,
  title={Text-Free Prosody-Aware Generative Spoken Language Modeling},
  author={Kharitonov, Eugene and Lee, Ann and Polyak, Adam and Adi, Yossi and Copet, Jade and Lakhotia, Kushal and Nguyen, Tu Anh and Riviere, Morgane and Mohamed, Abdelrahman and Dupoux, Emmanuel and others},
  booktitle={Proc. ACL},
  pages={8666--8681},
  year={2022}
}

@inproceedings{
tang2024salmonn,
title={{SALMONN}: Towards Generic Hearing Abilities for Large Language Models},
author={Changli Tang and Wenyi Yu and Guangzhi Sun and Xianzhao Chen and Tian Tan and Wei Li and Lu Lu and Zejun MA and Chao Zhang},
booktitle={The Twelfth International Conference on Learning Representations},
year={2024},
url={https://openreview.net/forum?id=14rn7HpKVk}
}

@inproceedings{gong2023joint,
  title={Joint audio and speech understanding},
  author={Gong, Yuan and Liu, Alexander H and Luo, Hongyin and Karlinsky, Leonid and Glass, James},
  booktitle={Proc. ASRU},
  pages={1--8},
  year={2023},
  organization={IEEE}
}

@article{chang2024speechprompt,
  title={Speechprompt: Prompting speech language models for speech processing tasks},
  author={Chang, Kai-Wei and Wu, Haibin and Wang, Yu-Kai and Wu, Yuan-Kuei and Shen, Hua and Tseng, Wei-Cheng and Kang, Iu-thing and Li, Shang-Wen and Lee, Hung-yi},
  journal={IEEE/ACM Transactions on Audio, Speech, and Language Processing},
  year={2024},
  publisher={IEEE}
}

@article{rubenstein2023audiopalm,
  title={Audio{PaLM}: A large language model that can speak and listen},
  author={Rubenstein, Paul K and Asawaroengchai, Chulayuth and Nguyen, Duc Dung and Bapna, Ankur and Borsos, Zal{\'a}n and Quitry, F{\'e}lix de Chaumont and Chen, Peter and Badawy, Dalia El and Han, Wei and Kharitonov, Eugene and others},
  journal={arXiv preprint arXiv:2306.12925},
  year={2023}
}

@article{chen2024vall,
  title={{VALL-E} 2: Neural Codec Language Models are Human Parity Zero-Shot Text to Speech Synthesizers},
  author={Chen, Sanyuan and Liu, Shujie and Zhou, Long and Liu, Yanqing and Tan, Xu and Li, Jinyu and Zhao, Sheng and Qian, Yao and Wei, Furu},
  journal={arXiv preprint arXiv:2406.05370},
  year={2024}
}

@article{wang2023neural,
  title={Neural codec language models are zero-shot text to speech synthesizers},
  author={Wang, Chengyi and Chen, Sanyuan and Wu, Yu and Zhang, Ziqiang and Zhou, Long and Liu, Shujie and Chen, Zhuo and Liu, Yanqing and Wang, Huaming and Li, Jinyu and others},
  journal={arXiv preprint arXiv:2301.02111},
  year={2023}
}

@inproceedings{zhang2023speechgpt,
  title={Speech{GPT}: Empowering Large Language Models with Intrinsic Cross-Modal Conversational Abilities},
  author={Zhang, Dong and Li, Shimin and Zhang, Xin and Zhan, Jun and Wang, Pengyu and Zhou, Yaqian and Qiu, Xipeng},
  booktitle={Findings of the Association for Computational Linguistics: EMNLP 2023},
  pages={15757--15773},
  year={2023}
}

@inproceedings{
kim2024clamtts,
title={{CL}aM-{TTS}: Improving Neural Codec Language Model for Zero-Shot Text-to-Speech},
author={Jaehyeon Kim and Keon Lee and Seungjun Chung and Jaewoong Cho},
booktitle={The Twelfth International Conference on Learning Representations},
year={2024},
url={https://openreview.net/forum?id=ofzeypWosV}
}

@inproceedings{maiti2024voxtlm,
  title={{VoxtLM}: Unified Decoder-Only Models for Consolidating Speech Recognition, Synthesis and Speech, Text Continuation Tasks},
  author={Maiti, Soumi and Peng, Yifan and Choi, Shukjae and Jung, Jee-weon and Chang, Xuankai and Watanabe, Shinji},
  booktitle={Proc. ICASSP},
  pages={13326--13330},
  year={2024},
  organization={IEEE}
}

@article{tian2024preference,
  title={Preference Alignment Improves Language Model-Based {TTS}},
  author={Tian, Jinchuan and Zhang, Chunlei and Shi, Jiatong and Zhang, Hao and Yu, Jianwei and Watanabe, Shinji and Yu, Dong},
  journal={arXiv preprint arXiv:2409.12403},
  year={2024}
}

@inproceedings{yanguniaudio,
  title={Uni{A}udio: Towards Universal Audio Generation with Large Language Models},
  author={Yang, Dongchao and Tian, Jinchuan and Tan, Xu and Huang, Rongjie and Liu, Songxiang and Guo, Haohan and Chang, Xuankai and Shi, Jiatong and Bian, Jiang and Zhao, Zhou and others},
  booktitle={Forty-first International Conference on Machine Learning},
  year={2024},
}

@article{Zeghidour2021SoundStreamAE,
  title={SoundStream: An End-to-End Neural Audio Codec},
  author={Neil Zeghidour and Alejandro Luebs and Ahmed Omran and Jan Skoglund and Marco Tagliasacchi},
  journal={IEEE/ACM Transactions on Audio, Speech, and Language Processing},
  year={2021},
  volume={30},
  pages={495-507},
  url={https://api.semanticscholar.org/CorpusID:236149944}
}

@article{Kumar2023HighFidelityAC,
  title={High-Fidelity Audio Compression with Improved RVQGAN},
  author={Rithesh Kumar and Prem Seetharaman and Alejandro Luebs and Ishaan Kumar and Kundan Kumar},
  journal={ArXiv},
  year={2023},
  volume={abs/2306.06546},
  url={https://api.semanticscholar.org/CorpusID:259138883}
}

@misc{shi2024espnetcodeccomprehensivetrainingevaluation,
      title={{ESPnet-Codec}: Comprehensive Training and Evaluation of Neural Codecs for Audio, Music, and Speech}, 
      author={Jiatong Shi and Jinchuan Tian and Yihan Wu and Jee-weon Jung and Jia Qi Yip and Yoshiki Masuyama and William Chen and Yuning Wu and Yuxun Tang and Massa Baali and others},
      year={2024},
      eprint={2409.15897},
      archivePrefix={arXiv},
      primaryClass={eess.AS},
      url={https://arxiv.org/abs/2409.15897}, 
}

@misc{wu2024speechcomposerunifyingmultiplespeech,
      title={Speech{C}omposer: Unifying Multiple Speech Tasks with Prompt Composition}, 
      author={Yihan Wu and Soumi Maiti and Yifan Peng and Wangyou Zhang and Chenda Li and Yuyue Wang and Xihua Wang and Shinji Watanabe and Ruihua Song},
      year={2024},
      eprint={2401.18045},
      archivePrefix={arXiv},
      primaryClass={cs.CL},
      url={https://arxiv.org/abs/2401.18045}, 
}

@inproceedings{zhang-etal-2023-speechgpt,
    title = "{S}peech{GPT}: Empowering Large Language Models with Intrinsic Cross-Modal Conversational Abilities",
    author = "Zhang, Dong  and
      Li, Shimin  and
      Zhang, Xin  and
      Zhan, Jun  and
      Wang, Pengyu  and
      Zhou, Yaqian  and
      Qiu, Xipeng",
    editor = "Bouamor, Houda  and
      Pino, Juan  and
      Bali, Kalika",
    booktitle = "Findings of the Association for Computational Linguistics: EMNLP 2023",
    month = dec,
    year = "2023",
    address = "Singapore",
    publisher = "Association for Computational Linguistics",
    url = "https://aclanthology.org/2023.findings-emnlp.1055",
    doi = "10.18653/v1/2023.findings-emnlp.1055",
    pages = "15757--15773",
}

@inproceedings{lee-etal-2022-direct,
    title = "Direct Speech-to-Speech Translation With Discrete Units",
    author = "Lee, Ann  and
      Chen, Peng-Jen  and
      Wang, Changhan  and
      Gu, Jiatao  and
      Popuri, Sravya  and
      Ma, Xutai  and
      Polyak, Adam  and
      Adi, Yossi  and
      He, Qing  and
      Tang, Yun  and
      Pino, Juan  and
      Hsu, Wei-Ning",
    booktitle = "Proc. ACL",
    year = "2022",
    publisher = "Association for Computational Linguistics",
    url = "https://aclanthology.org/2022.acl-long.235",
    doi = "10.18653/v1/2022.acl-long.235",
    pages = "3327--3339",

}

@inproceedings{NEURIPS2023_94b472a1,
 author = {Copet, Jade and Kreuk, Felix and Gat, Itai and Remez, Tal and Kant, David and Synnaeve, Gabriel and Adi, Yossi and Defossez, Alexandre},
 booktitle = {Advances in Neural Information Processing Systems},
 editor = {A. Oh and T. Naumann and A. Globerson and K. Saenko and M. Hardt and S. Levine},
 pages = {47704--47720},
 publisher = {Curran Associates, Inc.},
 title = {Simple and Controllable Music Generation},
 url = {https://proceedings.neurips.cc/paper_files/paper/2023/file/94b472a1842cd7c56dcb125fb2765fbd-Paper-Conference.pdf},
 volume = {36},
 year = {2023}
}

@inproceedings{Dhawan_2024,
  title     = {Codec-{ASR}: Training Performant Automatic Speech Recognition Systems with Discrete Speech Representations},
  author    = {Kunal Dhawan and Nithin Rao Koluguri and Ante Jukić and Ryan Langman and Jagadeesh Balam and Boris Ginsburg},
  year      = {2024},
  booktitle = {Interspeech},
  pages     = {2574--2578},
  doi       = {10.21437/Interspeech.2024-330},
  issn      = {2958-1796},
}

@misc{wu2024continuallearninglargelanguage,
      title={Continual Learning for Large Language Models: A Survey}, 
      author={Tongtong Wu and Linhao Luo and Yuan-Fang Li and Shirui Pan and Thuy-Trang Vu and Gholamreza Haffari},
      year={2024},
      eprint={2402.01364},
      archivePrefix={arXiv},
      primaryClass={cs.CL},
      url={https://arxiv.org/abs/2402.01364}, 
}

@misc{chu2024qwen2audiotechnicalreport,
      title={Qwen2-Audio Technical Report}, 
      author={Yunfei Chu and Jin Xu and Qian Yang and Haojie Wei and Xipin Wei and Zhifang Guo and Yichong Leng and Yuanjun Lv and Jinzheng He and Junyang Lin and Chang Zhou and Jingren Zhou},
      year={2024},
      eprint={2407.10759},
      archivePrefix={arXiv},
      primaryClass={eess.AS},
      url={https://arxiv.org/abs/2407.10759}, 
}

@misc{wu2024codecsuperbindepthanalysissound,
      title={Codec-{SUPERB}: An In-Depth Analysis of Sound Codec Models}, 
      author={Haibin Wu and Ho-Lam Chung and Yi-Cheng Lin and Yuan-Kuei Wu and Xuanjun Chen and Yu-Chi Pai and Hsiu-Hsuan Wang and Kai-Wei Chang and Alexander H. Liu and Hung-yi Lee},
      year={2024},
      eprint={2402.13071},
      archivePrefix={arXiv},
      primaryClass={eess.AS},
      url={https://arxiv.org/abs/2402.13071}, 
}

@misc{kim2024neuralspeechaudiocoding,
      title={Neural Speech and Audio Coding}, 
      author={Minje Kim and Jan Skoglund},
      year={2024},
      eprint={2408.06954},
      archivePrefix={arXiv},
      primaryClass={cs.SD},
      url={https://arxiv.org/abs/2408.06954}, 
}

@inproceedings{wang21s_interspeech,
  title     = {{CoVoST} 2 and Massively Multilingual Speech Translation},
  author    = {Changhan Wang and Anne Wu and Jiatao Gu and Juan Pino},
  year      = {2021},
  booktitle = {Interspeech},
  pages     = {2247--2251},
  doi       = {10.21437/Interspeech.2021-2027},
  issn      = {2958-1796},
}

@inproceedings{peng2023reproducing,
  title={Reproducing {W}hisper-style training using an open-source toolkit and publicly available data},
  author={Peng, Yifan and Tian, Jinchuan and Yan, Brian and Berrebbi, Dan and Chang, Xuankai and Li, Xinjian and Shi, Jiatong and Arora, Siddhant and Chen, William and Sharma, Roshan and others},
  booktitle={Proc. ASRU},
  pages={1--8},
  year={2023},
  organization={IEEE}
}

@inproceedings{bu2017aishell,
  title={Aishell-1: An open-source mandarin speech corpus and a speech recognition baseline},
  author={Bu, Hui and Du, Jiayu and Na, Xingyu and Wu, Bengu and Zheng, Hao},
  booktitle={Proc. O-COCOSDA},
  pages={1--5},
  year={2017},
  organization={IEEE}
}

@article{du2018aishell,
  title={Aishell-2: Transforming mandarin asr research into industrial scale},
  author={Du, Jiayu and Na, Xingyu and Liu, Xuechen and Bu, Hui},
  journal={arXiv preprint arXiv:1808.10583},
  year={2018}
}

@inproceedings{shi21c_interspeech,
  title     = {AISHELL-3: A Multi-Speaker Mandarin TTS Corpus},
  author    = {Yao Shi and Hui Bu and Xin Xu and Shaoji Zhang and Ming Li},
  year      = {2021},
  booktitle = {Interspeech},
  pages     = {2756--2760},
  doi       = {10.21437/Interspeech.2021-755},
  issn      = {2958-1796},
}

@inproceedings{zhang2022wenetspeech,
  title={Wenetspeech: A 10000+ hours multi-domain mandarin corpus for speech recognition},
  author={Zhang, Binbin and Lv, Hang and Guo, Pengcheng and Shao, Qijie and Yang, Chao and Xie, Lei and Xu, Xin and Bu, Hui and Chen, Xiaoyu and Zeng, Chenchen and others},
  booktitle={Proc. ICASSP},
  pages={6182--6186},
  year={2022},
  organization={IEEE}
}

@inproceedings{chen21o_interspeech,
  title     = {GigaSpeech: An Evolving, Multi-Domain {ASR} Corpus with 10,000 Hours of Transcribed Audio},
  author    = {Guoguo Chen and Shuzhou Chai and Guan-Bo Wang and Jiayu Du and Wei-Qiang Zhang and Chao Weng and Dan Su and Daniel Povey and Jan Trmal and Junbo Zhang and others},
  year      = {2021},
  booktitle = {Interspeech},
  pages     = {3670--3674},
  doi       = {10.21437/Interspeech.2021-1965},
  issn      = {2958-1796},
}

@inproceedings{panayotov2015librispeech,
  title={Librispeech: an asr corpus based on public domain audio books},
  author={Panayotov, Vassil and Chen, Guoguo and Povey, Daniel and Khudanpur, Sanjeev},
  booktitle={Proc. ICASSP},
  pages={5206--5210},
  year={2015},
  organization={IEEE}
}

@inproceedings{ardila-etal-2020-common,
    title = "Common Voice: A Massively-Multilingual Speech Corpus",
    author = "Ardila, Rosana  and
      Branson, Megan  and
      Davis, Kelly  and
      Kohler, Michael  and
      Meyer, Josh  and
      Henretty, Michael  and
      Morais, Reuben  and
      Saunders, Lindsay  and
      Tyers, Francis  and
      Weber, Gregor",
    booktitle = "Proc. LREC",
    month = may,
    year = "2020",
    address = "Marseille, France",
    publisher = "European Language Resources Association",
    url = "https://aclanthology.org/2020.lrec-1.520",
    pages = "4218--4222",
}

@article{rottland1997wsj,
  title={THE {WSJ} SPEECH DATABASE},
  author={Rottland, J and Neukirchen, Ch and Willett, D},
  year={1997}
}

@inproceedings{pratap20_interspeech,
  title     = {{MLS}: A Large-Scale Multilingual Dataset for Speech Research},
  author    = {Vineel Pratap and Qiantong Xu and Anuroop Sriram and Gabriel Synnaeve and Ronan Collobert},
  year      = {2020},
  booktitle = {Interspeech},
  pages     = {2757--2761},
  doi       = {10.21437/Interspeech.2020-2826},
  issn      = {2958-1796},
}

@inproceedings{hernandez2018ted,
  title={{TED-LIUM} 3: Twice as much data and corpus repartition for experiments on speaker adaptation},
  author={Hernandez, Fran{\c{c}}ois and Nguyen, Vincent and Ghannay, Sahar and Tomashenko, Natalia and Esteve, Yannick},
  booktitle={Speech and Computer: 20th International Conference, SPECOM 2018},
  pages={198--208},
  year={2018},
  organization={Springer}
}

@article{shoeybi2019megatron,
  title={Megatron-lm: Training multi-billion parameter language models using model parallelism},
  author={Shoeybi, Mohammad and Patwary, Mostofa and Puri, Raul and LeGresley, Patrick and Casper, Jared and Catanzaro, Bryan},
  journal={arXiv preprint arXiv:1909.08053},
  year={2019}
}

@article{korthikanti2023reducing,
  title={Reducing activation recomputation in large transformer models},
  author={Korthikanti, Vijay Anand and Casper, Jared and Lym, Sangkug and McAfee, Lawrence and Andersch, Michael and Shoeybi, Mohammad and Catanzaro, Bryan},
  journal={Proceedings of Machine Learning and Systems},
  volume={5},
  pages={341--353},
  year={2023}
}

@article{gao2020pile,
  title={The pile: An 800gb dataset of diverse text for language modeling},
  author={Gao, Leo and Biderman, Stella and Black, Sid and Golding, Laurence and Hoppe, Travis and Foster, Charles and Phang, Jason and He, Horace and Thite, Anish and Nabeshima, Noa and others},
  journal={arXiv preprint arXiv:2101.00027},
  year={2020}
}

@inproceedings{chang23b_interspeech,
  title     = {Exploration of Efficient End-to-End {ASR} using Discretized Input from Self-Supervised Learning},
  author    = {Xuankai Chang and Brian Yan and Yuya Fujita and Takashi Maekaku and Shinji Watanabe},
  year      = {2023},
  booktitle = {Interspeech},
  pages     = {1399--1403},
  doi       = {10.21437/Interspeech.2023-2051},
  issn      = {2958-1796},
}

@inproceedings{chang2024exploring,
  title={Exploring speech recognition, translation, and understanding with discrete speech units: A comparative study},
  author={Chang, Xuankai and Yan, Brian and Choi, Kwanghee and Jung, Jee-Weon and Lu, Yichen and Maiti, Soumi and Sharma, Roshan and Shi, Jiatong and Tian, Jinchuan and Watanabe, Shinji and others},
  booktitle={Proc. ICASSP},
  pages={11481--11485},
  year={2024},
  organization={IEEE}
}

@inproceedings{zen19_interspeech,
  title     = {Libri{TTS}: A Corpus Derived from LibriSpeech for Text-to-Speech},
  author    = {Heiga Zen and Viet Dang and Rob Clark and Yu Zhang and Ron J. Weiss and Ye Jia and Zhifeng Chen and Yonghui Wu},
  year      = {2019},
  booktitle = {Interspeech},
  pages     = {1526--1530},
  doi       = {10.21437/Interspeech.2019-2441},
  issn      = {2958-1796},
}

@inproceedings{radford2023robust,
  title={Robust speech recognition via large-scale weak supervision},
  author={Radford, Alec and Kim, Jong Wook and Xu, Tao and Brockman, Greg and McLeavey, Christine and Sutskever, Ilya},
  booktitle={International conference on machine learning},
  pages={28492--28518},
  year={2023},
  organization={PMLR}
}

@inproceedings{jung24c_interspeech,
  title     = {{ESP}net-{SPK}: full pipeline speaker embedding toolkit with reproducible recipes, self-supervised front-ends, and off-the-shelf models},
  author    = {Jee-Weon Jung and Wangyou Zhang and Jiatong Shi and Zakaria Aldeneh and Takuya Higuchi and Alex Gichamba and Barry-John Theobald and Ahmed {Hussen Abdelaziz} and Shinji Watanabe},
  year      = {2024},
  booktitle = {Interspeech},
  pages     = {4278--4282},
  doi       = {10.21437/Interspeech.2024-1345},
  issn      = {2958-1796},
}

@inproceedings{saeki2022utmos,
  title     = {{UTMOS}: UTokyo-SaruLab System for VoiceMOS Challenge 2022},
  author    = {Takaaki Saeki and Detai Xin and Wataru Nakata and Tomoki Koriyama and Shinnosuke Takamichi and Hiroshi Saruwatari},
  year      = {2022},
  booktitle = {Interspeech},
  pages     = {4521--4525},
  doi       = {10.21437/Interspeech.2022-439},
  issn      = {2958-1796},
}

@article{he2024emilia,
  title={Emilia: An extensive, multilingual, and diverse speech dataset for large-scale speech generation},
  author={He, Haorui and Shang, Zengqiang and Wang, Chaoren and Li, Xuyuan and Gu, Yicheng and Hua, Hua and Liu, Liwei and Yang, Chen and Li, Jiaqi and Shi, Peiyang and others},
  journal={arXiv preprint arXiv:2407.05361},
  year={2024}
}

@inproceedings{huang22f_interspeech,
  title     = {The {VoiceMOS} Challenge 2022},
  author    = {Wen Chin Huang and Erica Cooper and Yu Tsao and Hsin-Min Wang and Tomoki Toda and Junichi Yamagishi},
  year      = {2022},
  booktitle = {Interspeech},
  pages     = {4536--4540},
  doi       = {10.21437/Interspeech.2022-970},
  issn      = {2958-1796},
}

@inproceedings{ye23b_interspeech,
  title     = {GigaST: A 10,000-hour Pseudo Speech Translation Corpus},
  author    = {Rong Ye and Chengqi Zhao and Tom Ko and Chutong Meng and Tao Wang and Mingxuan Wang and Jun Cao},
  year      = {2023},
  booktitle = {Interspeech},
  pages     = {2168--2172},
  doi       = {10.21437/Interspeech.2023-1233},
  issn      = {2958-1796},
}

@inproceedings{post-2018-call,
    title = "A Call for Clarity in Reporting {BLEU} Scores",
    author = "Post, Matt",
    booktitle = "Proceedings of the Third Conference on Machine Translation: Research Papers",
    month = oct,
    year = "2018",
    address = "Brussels, Belgium",
    publisher = "Association for Computational Linguistics",
    url = "https://aclanthology.org/W18-6319",
    doi = "10.18653/v1/W18-6319",
    pages = "186--191",
}

@article{zhan2024anygpt,
  title={Any{GPT}: Unified multimodal {LLM} with discrete sequence modeling},
  author={Zhan, Jun and Dai, Junqi and Ye, Jiasheng and Zhou, Yunhua and Zhang, Dong and Liu, Zhigeng and Zhang, Xin and Yuan, Ruibin and Zhang, Ge and Li, Linyang and others},
  journal={arXiv preprint arXiv:2402.12226},
  year={2024}
}

@article{zhu2023vec,
  title={Vec-{S}ok speech: Speech vectorization and tokenization for neural speech generation},
  author={Zhu, Xinfa and Lv, Yuanjun and Lei, Yi and Li, Tao and He, Wendi and Zhou, Hongbin and Lu, Heng and Xie, Lei},
  journal={arXiv preprint arXiv:2310.07246},
  year={2023}
}

@article{huang2023speech,
  title={Speech translation with large language models: An industrial practice},
  author={Huang, Zhichao and Ye, Rong and Ko, Tom and Dong, Qianqian and Cheng, Shanbo and Wang, Mingxuan and Li, Hang},
  journal={arXiv preprint arXiv:2312.13585},
  year={2023}
}

@inproceedings{wu2024improving,
  title={Improving the Quality of {IWLST} 2024 Cascade Offline Speech Translation and Speech-to-Speech Translation via Translation Hypothesis Ensembling with NMT models and Large Language Models},
  author={Wu, Zhanglin and Guo, Jiaxin and Wei, Daimeng and Rao, Zhiqiang and Li, Zongyao and Shang, Hengchao and Luo, Yuanchang and Li, Shaojun and Yang, Hao},
  booktitle={Proc. IWSLT},
  pages={46--52},
  year={2024}
}

@article{defossez2024moshi,
  title={Moshi: a speech-text foundation model for real-time dialogue},
  author={D{\'e}fossez, Alexandre and Mazar{\'e}, Laurent and Orsini, Manu and Royer, Am{\'e}lie and P{\'e}rez, Patrick and J{\'e}gou, Herv{\'e} and Grave, Edouard and Zeghidour, Neil},
  journal={arXiv preprint arXiv:2410.00037},
  year={2024}
}

@inproceedings{
zhai2023investigating,
title={Investigating the Catastrophic Forgetting in Multimodal Large Language Model Fine-Tuning},
author={Yuexiang Zhai and Shengbang Tong and Xiao Li and Mu Cai and Qing Qu and Yong Jae Lee and Yi Ma},
booktitle={Conference on Parsimony and Learning (Proceedings Track)},
year={2023},
url={https://openreview.net/forum?id=g7rMSiNtmA}
}

@inproceedings{sun2020ernie,
  title={Ernie 2.0: A continual pre-training framework for language understanding},
  author={Sun, Yu and Wang, Shuohuan and Li, Yukun and Feng, Shikun and Tian, Hao and Wu, Hua and Wang, Haifeng},
  booktitle={Proc. AAAI},
  volume={34},
  number={05},
  pages={8968--8975},
  year={2020}
}

@inproceedings{paperno2016lambada,
  title={The {LAMBADA} dataset: Word prediction requiring a broad discourse context},
  author={Paperno, Denis and Kruszewski, Germ{\'a}n and Lazaridou, Angeliki and Pham, Ngoc-Quan and Bernardi, Raffaella and Pezzelle, Sandro and Baroni, Marco and Boleda, Gemma and Fern{\'a}ndez, Raquel},
  booktitle={Proc. ACL},
  pages={1525--1534},
  year={2016}
}

@article{xie2024mini,
  title={Mini-omni: Language models can hear, talk while thinking in streaming},
  author={Xie, Zhifei and Wu, Changqiao},
  journal={arXiv preprint arXiv:2408.16725},
  year={2024}
}

@article{copet2023simple,
  title={Simple and controllable music generation},
  author={Copet, Jade and Kreuk, Felix and Gat, Itai and Remez, Tal and Kant, David and Synnaeve, Gabriel and Adi, Yossi and D{\'e}fossez, Alexandre},
  journal={Advances in Neural Information Processing Systems},
  volume={36},
  pages={47704--47720},
  year={2023}
}

@inproceedings{zengscaling,
  title={Scaling Speech-Text Pre-training with Synthetic Interleaved Data},
  author={Zeng, Aohan and Du, Zhengxiao and Liu, Mingdao and Zhang, Lei and Dong, Yuxiao and Tang, Jie and others},
  booktitle={The Thirteenth International Conference on Learning Representations},
year={2025}
}

@inproceedings{wu2025towards,
  title={Towards Codec-{LM} Co-design for Neural Codec Language Models},
  author={Wu, Shih-Lun and Lahoti, Aakash and Desai, Arjun D and Goel, Karan and Donahue, Chris and Gu, Albert},
  booktitle={Proc. NAACL},
  pages={55--65},
  year={2025}
}

@inproceedings{casanova2025low,
  title={Low frame-rate speech codec: a codec designed for fast high-quality speech {LLM} training and inference},
  author={Casanova, Edresson and Langman, Ryan and Neekhara, Paarth and Hussain, Shehzeen and Li, Jason and Ghosh, Subhankar and Juki{\'c}, Ante and Lee, Sang-gil},
  booktitle={Proc. ICASSP},
  pages={1--5},
  year={2025},
  organization={IEEE}
}

@article{guo2024lscodec,
  title={{LSC}odec: Low-Bitrate and Speaker-Decoupled Discrete Speech Codec},
  author={Guo, Yiwei and Li, Zhihan and Du, Chenpeng and Wang, Hankun and Chen, Xie and Yu, Kai},
  journal={arXiv preprint arXiv:2410.15764},
  year={2024}
}

@inproceedings{ye2025codec,
  title={Codec does matter: Exploring the semantic shortcoming of codec for audio language model},
  author={Ye, Zhen and Sun, Peiwen and Lei, Jiahe and Lin, Hongzhan and Tan, Xu and Dai, Zheqi and Kong, Qiuqiang and Chen, Jianyi and Pan, Jiahao and Liu, Qifeng and others},
  booktitle={Proc. AAAI},
  volume={39},
  number={24},
  pages={25697--25705},
  year={2025}
}

@inproceedings{guo2025speaking,
  title={Speaking from coarse to fine: Improving neural codec language model via multi-scale speech coding and generation},
  author={Guo, Haohan and Xie, Fenglong and Yang, Dongchao and Wu, Xixin and Meng, Helen},
  booktitle={Proc. ICASSP},
  pages={1--5},
  year={2025},
  organization={IEEE}
}

@inproceedings{tian2025espnet,
  title={{ESPnet-SpeechLM}: An Open Speech Language Model Toolkit},
  author={Tian, Jinchuan and Shi, Jiatong and Chen, William and Arora, Siddhant and Masuyama, Yoshiki and Maekaku, Takashi and Wu, Yihan and Peng, Junyi and Bharadwaj, Shikhar and Zhao, Yiwen and others},
  booktitle={Proc. NAACL},
  pages={116--124},
  year={2025}
}

@article{xu2025qwen2,
  title={Qwen2. 5-omni technical report},
  author={Xu, Jin and Guo, Zhifang and He, Jinzheng and Hu, Hangrui and He, Ting and Bai, Shuai and Chen, Keqin and Wang, Jialin and Fan, Yang and Dang, Kai and others},
  journal={arXiv preprint arXiv:2503.20215},
  year={2025}
}

@article{ding2025kimi,
  title={Kimi-audio technical report},
  author={Ding, Ding and Ju, Zeqian and Leng, Yichong and Liu, Songxiang and Liu, Tong and Shang, Zeyu and Shen, Kai and Song, Wei and Tan, Xu and Tang, Heyi and others},
  journal={arXiv preprint arXiv:2504.18425},
  year={2025}
}

@article{barrault2023seamless,
  title={Seamless: Multilingual Expressive and Streaming Speech Translation},
  author={Barrault, Lo{\"\i}c and Chung, Yu-An and Meglioli, Mariano Coria and Dale, David and Dong, Ning and Duppenthaler, Mark and Duquenne, Paul-Ambroise and Ellis, Brian and Elsahar, Hady and Haaheim, Justin and others},
  journal={arXiv preprint arXiv:2312.05187},
  year={2023}
}

@inproceedings{agarwal2023findings,
  title={FINDINGS OF THE IWSLT 2023 EVALUATION CAMPAIGN},
  author={Agarwal, Milind and Agrawal, Sweta and Anastasopoulos, Antonios and Bentivogli, Luisa and Bojar, Ond{\v{r}}ej and Borg, Claudia and Carpuat, Marine and Cattoni, Roldano and Cettolo, Mauro and Chen, Mingda and others},
  booktitle={Proc. IWSLT},
  pages={1--61},
  year={2023}
}

@inproceedings{ahmad2024findings,
  title={FINDINGS OF THE IWSLT 2024 EVALUATION CAMPAIGN},
  author={Ahmad, Ibrahim Sa’id and Anastasopoulos, Antonios and Bojar, Ond{\v{r}}ej and Borg, Claudia and Carpuat, Marine and Cattoni, Roldano and Cettolo, Mauro and Chen, William and Dong, Qianqian and Federico, Marcello and others},
  booktitle={Proc. IWSLT},
  pages={1--11},
  year={2024}
}

\end{document}